\documentstyle[12pt,epsfig]{article}
  \advance\oddsidemargin  by -1in
  \advance\evensidemargin by -1in
\def\lsim{\mathrel{\rlap{\lower4pt\hbox{\hskip1pt$\sim$}}
    \raise1pt\hbox{$<$}}}         %less than or approx. symbol
\def\gsim{\mathrel{\rlap{\lower4pt\hbox{\hskip1pt$\sim$}}
    \raise1pt\hbox{$>$}}}         %greater than or approx. symbol

\def\be{\begin{equation}}
\def\ee{\end{equation}}
\def\bq{\begin{eqnarray}}
\def\eq{\end{eqnarray}}
\def\bm{\boldmath}
 
\mathchardef\alpha="710B
\mathchardef\beta="710C
\mathchardef\gamma="710D
\mathchardef\delta="710E
\mathchardef\epsilon="710F
\mathchardef\zeta="7110
\mathchardef\eta="7111
\mathchardef\theta="7112
\mathchardef\iota="7113
\mathchardef\kappa="7114
\mathchardef\lambda="7115
\mathchardef\mu="7116
\mathchardef\nu="7117
\mathchardef\xi="7118
\mathchardef\pi="7119
\mathchardef\rho="711A
\mathchardef\sigma="711B
\mathchardef\tau="711C
\mathchardef\upsilon="711D
\mathchardef\phi="711E
\mathchardef\chi="711F
\mathchardef\psi="7120
\mathchardef\omega="7121
\mathchardef\varepsilon="7122
\mathchardef\vartheta="7123
\mathchardef\varpi="7124
\mathchardef\varrho="7125
\mathchardef\varsigma="7126
\mathchardef\varphi="7127
\mathchardef\nabla="7272
 
\font\dozeb=cmmib10 scaled \magstep1
\font\dozesyb=cmbsy10 scaled \magstep1
\font\dezb=cmmib10
\def\bm{\fam9}
\begin{document}
\pagestyle{empty}

%\hfill{\large DFTT 68/96}
 
%\hfill{\large October 1996}
 
\vspace{2.0cm}
 
\begin{center}
 
{\Large \bf ON THE QCD EVOLUTION OF THE \\ TRANSVERSITY 
DISTRIBUTION \\}

\vspace{1.0cm}

{\large Vincenzo Barone\footnote{Also at 
II Facolt{\`a} di Scienze MFN, 15100 Alessandria, Italy.} \\}

\vspace{1.0cm} 

{\it $^{a}$Dipartimento di
Fisica Teorica, Universit\`a di Torino,\\
and INFN, Sezione di
Torino,    10125 Torino, Italy.\\} 

\vspace{1.0cm}

{\large \bf Abstract \bigskip\\ }

\end{center}

The QCD evolution of the 
transversity distributions is 
investigated and compared to that of the helicity 
distributions. It is shown that they differ largely in the small--$x$
region. It is also proved that the evolution preserves 
Soffer's
inequality among the three leading--twist distribution 
functions.

\vfill
 
\pagebreak

\baselineskip 24 pt
\pagestyle{plain}

The transversity distribution, originally introduced
by Ralston and Soper \cite{RS} (see also 
\cite{JJ,AM,CPR}), and now customarily called $h_1$, measures
the polarization asymmetry of quarks (or antiquarks) in a 
transversely polarized hadron. More explicitly, $h_1(x)$ is 
the number density of quarks polarized in a transverse
direction $+ \hat {\mbox{\bm $n$}}$ with a given longitudinal
momentum fraction 
$x$ minus the number density of quarks 
polarized in the opposite
direction $- \hat {\mbox{\bm $n$}}$, when the hadron's spin 
points in the direction $+ \hat {\mbox{\bm $n$}}$, {\it i.e.}
\be
h_1^q(x) = q_{+ \hat {\mbox{\bm $n$}}}(x) - 
q_{- \hat {\mbox{\bm $n$}}}(x) \,.
\label{eq1}
\ee 
In the Operator--Product--Expansion 
 language $h_1^q(x)$ is a leading twist quantity \cite{JJ}
and therefore
has the same status as the other  better known leading twist 
distribution functions, the unpolarized density $q(x)$ and the 
helicity distribution $\Delta q(x)$. However, $h_1^q$ is chirally
odd \cite{JJ} and decouples from inclusive deep inelastic scattering. 
This
makes it a rather elusive observable. Its measurement is possible
only in polarized hadron--hadron scattering or in semi-inclusive
reactions \cite{Ji,JJ2,JS}, 
the best method being probably the Drell--Yan 
dimuon production with two transversely polarized proton beams, an
experiment planned at  RHIC \cite{RHIC} (interest in this 
process has been also expressed in the HERA-$\vec N$ proposal
\cite{HERAN}).

Due to the lack of experimental data, our knowledge of the 
shape and magnitude of $h_1^q$
relies on model calculations \cite{BCD,Poby}. 
%, among which only one is actually 
%sophisticated enough to give us some realistic information 
%on $h_1^q$.  
An alternative adopted by 
some authors \cite{Ji,BS} to obtain predictions 
for measurable quantities related to $h_1$
is to use fits to the longitudinally polarized data with
the assumption $h_1^q \simeq \Delta q$ at all momentum scales.
% Now, since we know \cite{AM}
%that the leading order anomalous dimensions of $h_1^q$ and
%of $\Delta q$ are different (for more details, see below),
%the assumption of Refs.~\cite{Ji,BS} is that 
This means that 
the difference in the QCD evolution of $h_1^q$ and $\Delta q$  
in the $x$--space is considered to be irrelevant. 
We shall show that this is definitely 
not the case: even though
the first moments  $h_1^q(1,Q^2), \, 
\Delta q (1,Q^2)$ do not evolve very differently at first order
(actually $\Delta q (1,Q^2)$ is constant 
whereas $ h_1^q (1,Q^2)$ decreases very slowly with $Q^2$, 
due to the smallness
of its first anomalous dimension), the evolution in the shape
of the two distributions is dramatically different, especially
at small $x$, and should certainly be taken into account. A 
calculation of an important class of observables, the 
Drell--Yan double--spin transverse asymmetries, which 
correctly treats the evolution of $h_1^{q,\bar q}$ is 
contained in \cite{BCD2}.  

Another issue which deserves some attention is the 
inequality among the three leading twist distribution 
functions, $q, \Delta q$ and $h_1^q$, recently discovered
by Soffer \cite{Soffer}. 
The theoretical status of Soffer's inequality
is matter of discussion. This inequality was proved
in a parton model framework and it has  been argued \cite{GJJ}
that
it is spoiled by radiative corrections, much like 
the Callan--Gross relation. 
%Its 
%validity in the QCD improved parton model is thus uncertain. 
A question of interest is whether it is 
preserved by the QCD evolution. We shall answer positively
this question.

Let us start by looking in detail at the QCD 
evolution of the transverse polarization distribution. 
Being chirally odd, $h_1(x,Q^2)$ does not mix with gluon 
distributions, which are chirally even. Thus its $Q^2$
evolution at leading order is governed only by the 
process of gluon emission. The Altarelli--Parisi equation 
for the QCD evolution of $h_1(x,Q^2)$ at order $\alpha_s$
is ($t \equiv \log{\frac{Q^2}{\mu^2}}$)
\be
\frac{dh_1^{q,\bar q}(x,t)}{dt} = 
\frac{{\alpha}_s(t)}{2 \pi} \, 
\int_x^1 \frac{dz}{z} \, P_h(z) \, h_1^{q,\bar q}(\frac{x}{z}, t)\,,
\label{qcd1}
\ee
where the leading order splitting function $P_h(z)$ has been computed
by Artru and Mekhfi \cite{AM}  and reads
\be
P_h(z) = \frac{4}{3} \, \left [ \frac{2}{(1-z)_{+}} - 2 + 
\frac{3}{2} \, \delta (z-1) \right ]\,.
\label{qcd2}
\ee
Inserting (\ref{qcd2}) in eq.~(\ref{qcd1}) gives (for each quark
and antiquark flavor)
\bq
\frac{dh_1^{q,\bar q}(x,t)}{dt} &=& 
\frac{{\alpha}_s(t)}{2 \pi} \, \left \{ 
[ 2 + \frac{8}{3} \, \log {(1-x)} ] h_1(x,t) \right. 
\nonumber \\
&+& \left. \frac{8}{3} \, 
\int_x^1 \frac{dz}{z} \, 
\frac{1}{1-z} \, [h_1(\frac{x}{z}, t) - z \, h_1(x,t) ]
- \frac{8}{3} \, 
\int_x^1 \frac{dz}{z} \, h_1(\frac{x}{z}, t) \right \}\,.
\label{qcd3}
\eq
Notice that $P_h$ can be conveniently decomposed as 
\be
P_h(z) = P_{qq}(z) - \frac{4}{3} (1-z)\,,
\label{qcd31}
\ee
where $P_{qq}$ is the usual quark splitting function for gluon emission, 
that is 
\be
P_{qq}(z) = \frac{4}{3} \, \left [ \frac{1+z^2}{(1-z)_{+}}  + 
\frac{3}{2} \, \delta (z-1) \right ]\,.
\label{qcd32}
\ee
and the second term in the r.h.s. of (\ref{qcd32}), which we shall 
call $\delta P_h(z)$,  is responsible 
for the peculiar evolution of $h_1$.  Note that $\delta P_h$ 
is always negative. 
The decomposition (\ref{qcd31}) will prove useful in the following.

The 
$Q^2$ dependence of the moments of $h_1$, 
$h_1(N,Q^2) \equiv 
\int_0^1 \, dx \, x^{N-1} \, h_1(x,Q^2)$, 
is governed by the anomalous dimensions $\gamma_{N}^h$
({\it i.e.} the Mellin transforms of the splitting function $P_h(z)$), 
according to the 
multiplicative rule
\be
h_1(N,Q^2) = 
h_1(N,Q_0^2) \, \left [ 
\frac{\alpha_s(Q_0^2)}{\alpha_s(Q^2)} 
\right ]^{6 \gamma_{N}^h/(33-2 n_f)}\,,
\label{qcd4} 
\ee
where $n_f$ is the number of flavors. 
Explicitly, the leading order anomalous dimensions are
given by
\bq
\gamma_{N}^h &=& \frac{4}{3} \, \left (
\frac{3}{2} - 2 \, \sum_{j=1}^{N} \frac{1}{j} \right ) \nonumber \\
&=& \frac{4}{3} \, \left \{ \frac{3}{2} - 2 \, 
\left [ \psi(N+1) + \gamma_E \right ] \right \}
\label{qcd5}
\eq
where $\psi(z) \equiv  \frac{d \ln \Gamma(z)}{dz}$ 
is the digamma function and $\gamma_E$ is 
the Euler--Mascheroni constant. 
In particular, since $\gamma_1^h = -\frac{2}{3}$, 
the first moment of $h_1$ and the tensor charge 
$\delta q \equiv \int dx \, 
(h_1^q - h_1^{\bar q})$ decrease with $Q^2$ as
\be
\delta q(Q^2) = \delta q(Q_0^2) \, 
\left [ \frac{\alpha_s(Q_0^2)}{\alpha_s(Q^2)} 
\right ]^{-4/27}\,.
\label{qcd6}
\ee
The smallness of the exponent $-\frac{4}{27}$ might induce 
one to think that 
the evolution of $h_1^q(x,Q^2)$ is not much different from that 
of the helicity distributions $\Delta q(x,Q^2)$ (remember that 
the $q \rightarrow qG$
anomalous dimension $\gamma_1^{qq}$ vanishes at all orders and the 
$G \rightarrow q \bar q$ polarized anomalous dimension 
$\gamma_1^{qG}$ is zero at leading order, so that $\Delta q(1,Q^2)$
is constant). 
As a matter of fact, the evolution in the $x$--space is 
sensibly different, especially at small $x$. 

This can be seen analytically by an argument 
based on the double--log approximation.  
 The leading behavior 
of the parton distributions at small $x$ is governed by the 
rightmost singularity of their  anomalous dimensions in the $N$--space. 
For $h_1^q$ eq.~(\ref{qcd5}) shows that this singularity is located 
at $N=-1$. Expanding $\gamma_N^h$ around this point gives
\be
\gamma_N^h \sim - \frac{1}{N+1} + {\cal O}(1)
\label{qcd7}
\ee
In the $x$--space,  
expanding the splitting function $P_h$ in powers of $x$ yields
\be
P_h(x) \sim \frac{8}{3} x + {\cal O}(x^2)\,.
\label{qcd8}
\ee
By contrast, in the longitudinally polarized case
it is known \cite{F} that for 
$\Delta q$ 
the rightmost singularity in the space of moments 
is located at $N=0$ and the splitting functions $\Delta P_{qq}$ and
$\Delta P_{qg}$ behave as constants as $x \rightarrow 0$. 
This means that, in the QCD evolution at small $x$,  
$h_1^q$ is suppressed by a power of $x$ with respect to $\Delta q$.

We can investigate numerically this problem by
solving the Altarelli--Parisi 
equation (\ref{qcd1}) with a suitable input for $h_1$. 
We assume $h_1^q$ and $\Delta q$ to be equal 
at a small scale $Q_0^2$ and 
let the two distributions evolve differently, according to their
own evolution equation. The assumption $h_1^q(x,Q_0^2) = 
\Delta q(x,Q_0^2)$
 is suggested by
quark model calculations \cite{JJ,BCD} 
of $h_1^q$ and $\Delta q$, which show
that these two 
distributions  are almost equal at a scale $Q_0^2 \lsim 0.5$ GeV$^2$.
For $\Delta q(x,Q_0^2)$ we use the leading order GRV parametrization 
\cite{GRV}  
whose input scale is $Q_0^2 = 0.23$ GeV$^2$. 
The result for the $u$ 
distributions is shown in Fig.~1 (the situation is similar
for the other flavors). 
The dashed line is the input, the solid line and the dotted 
line are the results of the evolution of $h_1^u$ and $\Delta u$, 
respectively, at $Q^2 = 25$ GeV$^2$. For completeness the evolution
of $h_1^u$ driven only by $P_{qq}$, 
with the $\delta P_h$ term turned off -- 
see eq.~(\ref{qcd31}) -- 
is also shown (dot-dashed line). 
The large 
difference in the evolution of $h_1^u$ (solid curve)
and $\Delta u$ (dotted curve) at small $x$ 
is evident. Notice also the discrepancy between the correct 
evolution of $h_1^u$ and the evolution driven 
by $P_{qq}$ (dot-dashed curve). 
  
Let us come now to Soffer's 
inequality among the three leading twist distribution 
functions. It reads \cite{Soffer}
\be
q(x) + \Delta q(x) \ge 2 \, \vert h_1^q(x) \vert\,,
\label{qcd9}
\ee
or, equivalently, $q_{+}(x) \ge \vert h_1^q(x) \vert$, having 
introduced  $q_{\pm} \equiv \frac{1}{2}(q \pm \Delta q)$. 
This relation has been rigorously proved by 
Soffer in the parton model, 
by relying on a positivity bound on the quark--nucleon 
forward amplitudes. A rederivation of the inequality was offered in 
\cite{GJJ}, where it was pointed out that it is spoiled by radiative
corrections and it was claimed that 
its status is similar to that of the 
Callan--Gross relation. 
The question arises whether Soffer's inequality is preserved 
by the QCD evolution (at least at leading order, which is all we 
know at present). We shall show that the answer to this question is 
positive and can be obtained in a very simple manner. 

Explicitly stated, the problem is: assuming (\ref{qcd9}) to be valid at 
some scale, will it hold at any larger scale ? In order to prove that it 
is indeed so, it suffices to show that the rate of evolution 
of $\vert h_1^q \vert$  is always smaller than that of $q_+$, namely
\be
\frac{d \vert h_1^q \vert}{d t} \le
\frac{d q_+}{dt} \,.
\label{qcd10}
\ee

Now, the Altarelli--Parisi equation for $q_+$ is ($\otimes$ denotes
convolution) 
%that is $A \otimes B \equiv \int_0^1 dz/z \, A(z) \, 
%B(x/z)$)
\be
\frac{dq_+}{dt} = \frac{\alpha_s(t)}{2 \pi} \, \left 
( P_{qq} \otimes q_+  + P_{qG}^{(+)} \otimes G_+ + 
P_{qG}^{(-)} \otimes G_- \right )\,,
\label{qcd11}
\ee
where $G_{\pm}(x,Q^2)$ are the gluon helicity distributions,  
and the two $G \rightarrow q \bar q$ splitting 
functions are $P_{qG}^{(+)}(z) = \frac{1}{2}z^2$ and $P_{qG}^{(-)}(z) = 
\frac{1}{2}(1-z)^2$.
  
Comparing the evolution equations (\ref{qcd11}) and (\ref{qcd1}) 
(the latter written for $\vert h_1^q \vert$), 
the inequality  (\ref{qcd10}) follows immediately from 
two facts: 
\begin{enumerate}
\item
The splitting functions $P_{qG}^{(\pm)}(z)$ are positive definite.

\item
 $P_h$ always gives a contribution to the 
convolution integral smaller than that of 
$P_{qq}$, since $\delta P_h$ is negative, see eq.~(\ref{qcd31}). 

\end{enumerate}

Therefore, Soffer's inequality is not spoiled by the 
QCD evolution, in much the same way as the positivity
bound $\vert \Delta q \vert \le q$ is protected, because
the evolution can never make the probabilities
$q_{\pm}$ become negative.  
Hence, it is with this positivity relation  that an analogy can be 
made, rather than with the Callan--Gross relation. 
The latter  
involves structure functions, {\it i.e.} physical quantities, 
whereas Soffer's inequality is a relation among distribution 
functions. The reason why eq.~(\ref{qcd9})  
has eluded for such a longtime the attention of physicists
 is simply that it does not have a probabilistic interpretation, 
for it involves nondiagonal quark--nucleon 
amplitudes. 

In conclusion, let us summarize our results. First of all, we have 
shown that the QCD evolution of $h_1(x,Q^2)$ possesses some 
relevant peculiarities which make it largely different from 
that of the helicity distributions at small $x$. This difference
can by no means be neglected if one wants to make reliable
predictions on experimentally accessible quantities (especially
when these
 are sensitive to the small-$x$ region,
as it is  the case of some double--spin
 transverse asymmetries \cite{BCD2})
Second, we have shown that Soffer's inequality is 
protected 
by the QCD evolution and thus it is precisely on the same ground 
as the positivity relation between polarized and unpolarized 
distribution functions.

\vspace{0.5cm}
I am grateful to M.~Anselmino,
 T.~Calarco, A.~Drago, S.~Forte and X.~Ji for useful 
discussions. 
I would also like to thank the Institute of Nuclear Theory
at the University of Washington for its hospitality and the U.S.
Department of Energy for partial support during 
an early stage of this work.

\pagebreak

\pagebreak

\begin{figure}[ht]
\mbox{\epsfig{file=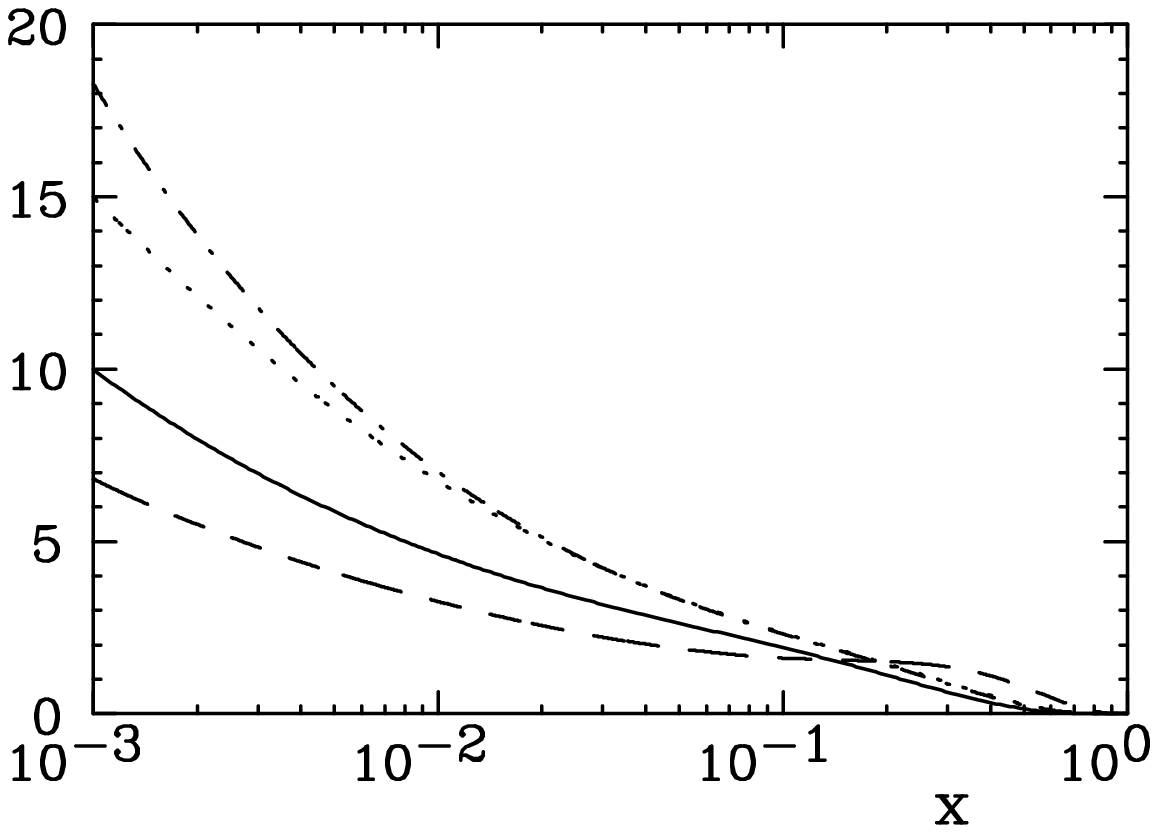,width=0.8\textwidth}}
\caption{Evolution of the helicity and transversity distributions 
for the $u$ flavor.
The dashed curve is the input $h_1^u \equiv \Delta u$ at 
$Q_0^2 =0.23$ GeV$^2$ taken from the GRV [17] parametrization. 
The solid (dotted) curve is $h_1^u$ ($\Delta u$) at $Q^2 = 25$ GeV$^2$. 
The dot-dashed curve 
is the result of the evolution of 
$h_1^u$ at $Q^2 =25$ GeV$^2$ driven by $P_{qq}$, {\it i.e.} with the 
 term $\delta P_h$ turned off in $P_h$.}
\end{figure}

\end{document}